\documentclass[conference]{IEEEtran}
\IEEEoverridecommandlockouts
\usepackage{cite}
\usepackage{amsmath,amssymb,amsfonts}
\usepackage{algorithmic}
\usepackage{graphicx}
\usepackage{textcomp}
\usepackage{xcolor}
\usepackage{comment}
\usepackage{url}
\usepackage{booktabs}
\usepackage{multirow}
\def\BibTeX{{\rm B\kern-.05em{\sc i\kern-.025em b}\kern-.08em
    T\kern-.1667em\lower.7ex\hbox{E}\kern-.125emX}}
  
\begin{document}

\title{RAG Strategies for Natural Language-Based\\SQL Query and REST API Call Generation}

\author{\IEEEauthorblockN{Tim Schlippe, Simon Martin, Michael Marketsmüller}
\IEEEauthorblockA{
\textit{IU International University of Applied Sciences}\\
Germany \\
tim.schlippe@iu.org}
}

\maketitle

\begin{abstract}
Enterprise software systems commonly expose business functionality through both relational databases and REST APIs. Accessing these interfaces typically requires specialized technical knowledge, as users must determine whether a request requires a database query or an API operation and understand the corresponding schemas, endpoints, and parameters. This creates demand for natural language interfaces that can translate user requests into SQL queries and REST API calls. 
While large language models (LLMs) show promise for structured code generation \cite{chen2021evaluating,huynh2025survey}, they typically lack reliable knowledge of enterprise-specific schemas, endpoints, and documentation. Retrieval-augmented generation (RAG) addresses this limitation by grounding generation in external documentation. However, prior work largely studies SQL query generation and REST API call generation as separate tasks, despite enterprise documentation environments often containing both database schemas and API specifications simultaneously. 
This paper systematically evaluates three training-free RAG configurations---\textit{Standard RAG}, \textit{Relevance Filtering RAG}, and \textit{Iterative Retrieval RAG}---across SQL query generation, REST API call generation, and a combined task requiring routing between both operation types. Using SAP Transactional Banking as a realistic enterprise use case, we constructed an expert-reviewed dataset with mock execution screening and compared retrieval strategies under database-only, API-only, and mixed-documentation settings. 
The results show that retrieval augmentation substantially improves structured generation performance in the controlled evaluation: Without retrieval, exact match accuracy is 0\% across all tasks, whereas retrieval yields mock execution success of up to 79.30\% and component match accuracy of up to 78.86\%. In the combined SQL query and REST API call setting, \textit{Iterative Retrieval RAG} achieves higher exact match accuracy than \textit{Standard RAG} (10.29\% vs.\ 7.45\%), while the difference in mock execution success is smaller (68.93\% vs.\ 67.10\%). For REST API call generation under mixed documentation, however, \textit{Iterative Retrieval RAG} achieves lower mock execution success than \textit{Standard RAG} (50.53\% vs.\ 75.79\%). Overall, the findings show that retrieval-control strategies exhibit task- and metric-dependent behavior under mixed-documentation settings.
\end{abstract}

\begin{IEEEkeywords}
Retrieval-Augmented Generation, LLMs, Large Language Models, Text-to-SQL, REST API Call Generation
\end{IEEEkeywords}

\section{Introduction}

Enterprise software systems increasingly expose functionality through two complementary interfaces: databases for ad-hoc information retrieval via SQL and REST APIs for transactional operations and data modification \cite{koutrika2024natural,wang2024fantastic}. Effective use of these interfaces requires specialized technical knowledge: Business users must understand SQL syntax and database schemas, while invoking APIs requires knowledge of endpoint specifications, HTTP methods, and payload structures \cite{usta2021dbtagger}. This technical barrier creates operational bottlenecks because domain experts often depend on IT specialists for routine data access and system interaction tasks \cite{lennerholt2020user}.

Recent advances in large language models (LLMs) offer a promising way to translate natural language requests into structured outputs such as SQL queries and REST API calls \cite{chen2021evaluating,huynh2025survey}. However, enterprise deployment remains challenging since LLMs typically lack reliable knowledge of domain-specific schemas, proprietary APIs, and evolving enterprise documentation \cite{chen2025llms}. This knowledge gap can lead to hallucinations \cite{ryser2025calibrated}, resulting in syntactically plausible but semantically incorrect outputs that reference non-existent tables, columns, endpoints, or parameters.

Retrieval-augmented generation (RAG) addresses this limitation by incorporating external documentation into the generation process \cite{lewis2020retrieval}. By retrieving relevant schema definitions, API specifications, or usage examples at inference time, RAG systems can ground generated outputs in available system documentation. Prior work has shown that documentation retrieval can improve code generation for unfamiliar libraries and APIs \cite{chen2025llms,zhou2023docprompting,kivroglou2025investigating}. These findings suggest that retrieval is particularly relevant for enterprise settings, where correctness depends less on general programming knowledge than on access to system-specific documentation.

Despite these advances, current research leaves three gaps: First, existing evaluations typically study text-to-SQL \cite{yu2019spider,lei2024evaluating} and REST API call generation \cite{kim2024seal} as separate tasks, with distinct datasets, metrics, and assumptions. In practical enterprise assistants, however, user requests may require the system to decide whether a request should be answered through a database query or through an API call. This creates a combined task in which routing and generation must be evaluated together. Second, the behavior of RAG approaches under mixed-documentation settings remains insufficiently understood. When database schemas and API specifications are available within the same retrieval environment, semantically related but task-irrelevant documentation may introduce retrieval noise. Third, existing work provides limited evidence on whether training-free adaptations of retrieval-control principles from Self-RAG \cite{asai2024selfrag} and CoRAG \cite{wang2025chain}, such as relevance filtering and iterative retrieval, offer advantages for structured generation tasks beyond standard top-$k$ retrieval.

This paper investigates these questions in the context of SAP Transactional Banking. We evaluate \textit{Standard RAG}, \textit{Relevance Filtering RAG}, and \textit{Iterative Retrieval RAG} across SQL query generation, REST API call generation, and a combined task requiring routing between both operation types. The latter two configurations provide training-free adaptations of retrieval-control principles from Self-RAG and CoRAG, respectively, while keeping the API-based GPT-5 backbone fixed across all experimental conditions. The evaluation uses database-only, API-only, and mixed-documentation settings to analyze how retrieval strategies behave when documentation sources vary in scope and heterogeneity.

This paper makes the following contributions:
\begin{itemize}
    \item \textbf{Joint SQL+API evaluation setting:} We evaluate natural language-based SQL query generation, REST API call generation, and a combined task requiring routing between both operation types within a shared enterprise retrieval environment.
    
    \item \textbf{Adaptable dataset construction pipeline:} We construct a test set of 631 cases (346 SQL, 285 API) grounded in SAP Transactional Banking documentation, screened through mock execution, and independently reviewed by three reviewers. The pipeline is adaptable to other enterprise schemas and API documentation, and the dataset is publicly available\footnote{https://github.com/mmiustudy/RAG4SQL-API}
    
    
    \item \textbf{Retrieval strategy analysis:} We compare \textit{Standard RAG}, \textit{Relevance Filtering RAG}, and \textit{Iterative Retrieval RAG} under database-only, API-only, and mixed-documentation settings, identifying how training-free retrieval-control strategies affect structured generation performance when documentation sources become heterogeneous.
\end{itemize}

The remainder of this paper is organized as follows: Section~\ref{sec:relatedwork} reviews text-to-SQL generation, REST API call generation with LLMs, and RAG approaches. Section~\ref{sec:methodology} describes the experimental setup, including RAG configuration selection, dataset construction, documentation embedding, and evaluation metrics. Section~\ref{sec:results} presents results by task type, RAG configuration, and documentation setting. Section~\ref{sec:discussion} discusses implications for retrieval-based enterprise assistants, including retrieval necessity, mixed-documentation settings, task-specific retrieval behavior, and deployment considerations. Section~\ref{sec:conclusion} summarizes the key findings and outlines future directions.

\section{Related Work}
\label{sec:relatedwork}

This section reviews prior work on text-to-SQL generation, REST API call generation with LLMs, and retrieval-augmented generation. We focus on how existing work treats these tasks separately and why mixed-documentation settings remain insufficiently studied.

\subsection{Text-to-SQL Generation}

Natural language interfaces to databases have been studied extensively, with modern approaches leveraging neural semantic parsing \cite{yu2019spider}. The Spider benchmark established cross-domain evaluation with 200 databases and more than 10,000 questions, introducing component match and execution accuracy metrics \cite{yu2019spider}. Subsequent work explored conversational SQL generation through CoSQL \cite{yu2019cosql} and SParC \cite{yu2019sparc}, as well as more realistic enterprise-oriented workflows in Spider 2.0 \cite{lei2024evaluating}. Recent approaches incorporate retrieval by using schema-linking examples \cite{pourreza2023evaluating} or documentation \cite{zhou2023docprompting}. However, these studies focus on SQL query generation as a standalone task and do not evaluate how SQL generation behaves when database documentation is retrieved together with API documentation.

\subsection{REST API Call Generation with LLMs}

Tool-augmented LLMs extend model capabilities through external API invocation \cite{schick2023toolformer}. API-Bank \cite{li2023apibank} introduced a benchmark for evaluating multi-turn API usage in dialogue systems, while SEAL \cite{kim2024seal} provides structured evaluation through simulation across endpoint selection, parameter filling, and response generation. ToolBench \cite{qin2023tool} and Gorilla \cite{patil2023gorilla} demonstrate LLM fine-tuning for API documentation comprehension. Recent work further addresses nested API sequences \cite{basu2025nestful} and live code execution \cite{jain2024livecode}. These efforts substantially advance API-related tool use, but they remain focused on REST API call generation and do not consider settings in which the LLM must route between API calls and SQL queries within the same enterprise domain.

\subsection{Retrieval-Augmented Generation}

RAG~\cite{lewis2020retrieval} improves LLM outputs by conditioning generation on documents retrieved from external knowledge bases. Self-RAG \cite{asai2024selfrag} trains a model to control retrieval and critique generated outputs using reflection tokens. CoRAG \cite{wang2025chain} trains retrieval chains and performs iterative retrieval through query decomposition and dedicated test-time decoding strategies. Most evaluations of these approaches focus on question answering, reasoning, or dialogue tasks. Their underlying retrieval-control principles have received limited attention in structured generation tasks such as SQL query generation and REST API call generation.

For code generation, prior work has shown that retrieving documentation can improve performance when LLMs use unfamiliar libraries or APIs \cite{zhou2023docprompting,chen2025llms,kivroglou2025investigating}. However, these studies typically rely on standard retrieval pipelines and provide limited evidence on training-free retrieval-control strategies based on relevance filtering or iterative retrieval. Moreover, they generally evaluate retrieval over a single documentation type, leaving open how retrieval strategies behave when database schemas and API specifications are available in the same retrieval environment.

\subsection{Research Gaps}

Existing work provides strong foundations for SQL query generation, REST API call generation, and retrieval-augmented generation. However, three gaps remain: First, SQL query generation and REST API call generation are usually evaluated as separate tasks, which prevents analysis of combined tasks where a system must route between both operation types. Second, prior work provides limited evidence on how retrieval strategies behave under mixed-documentation settings containing both database schemas and API specifications. Third, training-free retrieval-control strategies based on relevance filtering and iterative retrieval have not been systematically compared with standard RAG for natural language-based structured generation in a shared enterprise domain. Our work addresses these gaps through a controlled evaluation across SQL query generation, REST API call generation, and a combined task using SAP Transactional Banking documentation.

\section{Methodology}
\label{sec:methodology}

This section describes the experimental setup for evaluating retrieval strategies for SQL query generation and REST API call generation. We first outline the overall evaluation design and implementation of each retrieval configuration, then describe the SAP Transactional Banking use case, dataset construction, documentation embedding pipeline, and evaluation metrics.

\subsection{Experimental Design}

We evaluated three training-free RAG configurations---\textit{Standard RAG}, \textit{Relevance Filtering RAG}, and \textit{Iterative Retrieval RAG}---together with a no-retrieval baseline across three task settings: (1) SQL query generation, (2) REST API call generation, and (3) a combined task in which the system must first determine whether a user request requires a database query or an API call before generating the corresponding output. Retrieval settings were selected according to task relevance: SQL query generation was evaluated under database-only and mixed-documentation settings, REST API call generation under API-only and mixed-documentation settings, and the combined task under mixed-documentation settings.

All configurations shared the same underlying infrastructure. We used OpenAI's \texttt{text-embedding-3-small} model (1,536 dimensions) for embedding generation, ChromaDB for vector storage using squared Euclidean distance similarity, and GPT-5 as the backbone LLM. All experiments used GPT-5 with the default decoding configuration provided by the OpenAI API. Each retrieval step returned the top-5 most similar documentation chunks, following prior work on retrieval-augmented code generation \cite{chen2025llms,zhou2023docprompting}. All experiments followed a single-turn prompt setup with separate developer and user instructions. Generated outputs were stored in JSON format and evaluated offline using the same automated evaluation pipeline across all retrieval configurations. 
The evaluated retrieval configurations were implemented as follows:

\begin{itemize}
    \item \textbf{\textit{Standard RAG}:} The system embeds the user request, retrieves the five most similar documentation chunks, concatenates the retrieved context with task instructions, and generates the final output in a single LLM call.
    
    \item \textbf{\textit{Relevance Filtering RAG}:} After initial retrieval, the LLM evaluates the relevance of each retrieved chunk individually using a relevance threshold of 0.2. Only chunks classified as relevant are retained for final prompt construction.
    
    \item \textbf{\textit{Iterative Retrieval RAG}:} The system performs prompted iterative retrieval through query decomposition. A sub-query is first generated from the user request, relevant documentation is retrieved for that sub-query, and the LLM then determines whether additional retrieval is required before generating the final output.
\end{itemize}

The latter two configurations adapt retrieval-control principles explored in Self-RAG \cite{asai2024selfrag} and CoRAG \cite{wang2025chain}, respectively, but are not faithful implementations of the original frameworks. \textit{Relevance Filtering RAG} uses post-retrieval LLM-based relevance assessment without Self-RAG's trained reflection-token mechanism, while \textit{Iterative Retrieval RAG} uses prompted query decomposition and repeated retrieval without CoRAG's trained retrieval chains and dedicated decoding procedure. This training-free design keeps the GPT-5 backbone unchanged across configurations, allowing the evaluation to focus on differences in retrieval control rather than model-specific training effects.

All prompts enforced strict task constraints. SQL query generation was restricted to \texttt{SELECT} statements only, while REST API call generation excluded \texttt{GET} requests and focused on transactional operations using \texttt{POST}, \texttt{PUT}, \texttt{PATCH}, and \texttt{DELETE}. Prompts additionally specified structured JSON output formatting and included a fixed reference date (May 1, 2025) to ensure temporal consistency.

\subsection{SAP Transactional Banking Use Case}

We used SAP Transactional Banking (TRBK) \cite{sap2025trbk} as the enterprise domain for evaluation. TRBK was selected for three reasons: First, it provides publicly accessible, well-structured OpenAPI specifications for multiple banking business objects, including accounts, cards, and loans. Second, it represents a realistic enterprise environment with domain-specific terminology and complex structured documentation. Third, although no public database schema is available, the API specifications enabled systematic derivation of aligned relational schemas.

A custom script extracted relational database schemas from \texttt{GET} endpoint response definitions in the OpenAPI specifications. Objects containing at least two meaningful columns were retained. The resulting schemas included table names, column definitions, data types, and descriptions, yielding 22 database tables aligned with 20 API business objects.

\subsection{Dataset Construction}

As no existing benchmark jointly evaluates SQL query generation and REST API call generation within a shared enterprise setting, we constructed a custom dataset using an adapted APIGen pipeline \cite{liu2024apigen}. The dataset construction process was designed to be adaptable to other enterprise environments as test-case generation is grounded in enterprise-specific documentation artifacts, including database schemas, API specifications, and semantic business object descriptions.  Dataset construction consisted of four stages:

\begin{enumerate}
    \item \textbf{Automated generation:} GPT-5 generated candidate test cases for each business object using the corresponding API specification or database schema together with semantic business object descriptions. Prompts instructed the LLM to generate natural language requests paired with reference SQL queries or REST API calls.

    \item \textbf{Input humanization:} A second GPT-5 pass rewrote generated requests into more conversational natural language while preserving factual constraints such as identifiers, dates, and parameters.

    \item \textbf{Mock execution screening:} All generated outputs were screened for successful execution in the corresponding mock environment. SQL queries were parsed using \texttt{sqlparse} and executed against a mock SQLite database populated with TRBK-derived schemas. REST API calls were executed against Postman mock servers \cite{postman2025} implementing the corresponding TRBK endpoints. Successful API execution required server responses with status codes in \{200, 201, 204\}.

    \item \textbf{Expert review:} Three reviewers with software engineering and banking-domain experience independently reviewed all candidate test cases for correctness, consistency, and alignment between natural language requests and reference outputs. If any reviewer rejected a test case, the case was removed from the dataset rather than corrected or adjudicated.
\end{enumerate}

The final dataset contains 631 test cases comprising 346 SQL cases and 285 REST API cases. 

\subsection{Documentation Embedding}

To support retrieval for both task types, we processed three forms of TRBK documentation into semantic chunks:

\begin{itemize}
    \item \textbf{Database schema documentation:} Each database schema containing table definitions, column names, types, and descriptions formed one semantic chunk. These chunks supported SQL query generation.

    \item \textbf{API documentation:} Each API endpoint definition containing supported HTTP methods, parameters, and response schemas formed one semantic chunk. Chunks exceeding 8,000 tokens were split using fixed-length chunking with 800-token overlap following prior retrieval studies \cite{bhat2025rethinking,amiri2024chunk}. These chunks supported REST API call generation.

    \item \textbf{Business object descriptions:} High-level semantic descriptions of business entities from SAP documentation were included to provide shared domain context across both task types.
\end{itemize}

This process yielded 218 documentation chunks in total, consisting of 22 database schema chunks, 174 API documentation chunks, and 22 business object description chunks. All chunks were embedded using \texttt{text-embedding-3-small} and stored in separate ChromaDB collections. Database-only experiments retrieved schema and business object chunks, API-only experiments retrieved API and business object chunks, and mixed-documentation experiments retrieved all chunk types together.

\subsection{Evaluation Metrics}

Following prior work on text-to-SQL and API evaluation \cite{yu2019spider,pourreza2023evaluating,kim2024seal,basu2025nestful}, we reported five evaluation metrics:

\begin{itemize}
    \item \textbf{Exact Match Accuracy:} Binary correctness after structural normalization. SQL evaluation compared clause-level equivalence, while REST API evaluation required matching HTTP method, endpoint, and request body.

    \item \textbf{Component Match Accuracy:} Partial correctness for individual output components. SQL evaluation considered clause-level correctness, while API evaluation considered method, endpoint, and parameter correctness.

    \item \textbf{Mock Execution Success:} Percentage of generated outputs that executed successfully against the mock database or mock API servers. This metric measures execution validity in the controlled mock environment and does not by itself establish semantic correctness.

    \item \textbf{Endpoint Retrieval Accuracy (API only):} Percentage of API cases in which the correct endpoint was selected independent of parameter correctness.

    \item \textbf{Classification Accuracy (combined task only):} Percentage of cases in which the LLM correctly routed the request to SQL query generation or REST API call generation.
\end{itemize}

Statistical significance for exact match accuracy was assessed using paired two-tailed $t$-tests with a significance threshold of $\alpha = 0.05$. Exact match was treated as a binary per-test-case outcome, and we report $p$-values together with 95\% confidence intervals for paired mean differences where relevant.

\section{Results}
\label{sec:results}

This section presents the experimental results for REST API call generation, SQL query generation, and the combined task. We compare \textit{Standard RAG}, \textit{Relevance Filtering RAG}, and \textit{Iterative Retrieval RAG} against a no-retrieval baseline under database-only, API-only, and mixed-documentation settings.

We report exact match accuracy (EM), component match accuracy (CM), mock execution success (Mock Exec.), endpoint retrieval accuracy (ER), and classification accuracy (Class) depending on the evaluated task setting.

\subsection{REST API Call Generation}

Table \ref{tab:api} summarizes the results for REST API call generation. Without retrieval augmentation, the baseline achieves 0\% exact match accuracy and 0\% endpoint retrieval accuracy. Although the baseline reaches 28.45\% mock execution success, the 0\% endpoint retrieval accuracy demonstrates that successful execution in the mock environment does not by itself establish semantic correctness.

\begin{table}[h!]
\centering
\caption{REST API call generation results (\%).}
\footnotesize
\setlength{\tabcolsep}{2.5pt}
\begin{tabular}{llrrrr}
\toprule
\textbf{Setting} & \textbf{Approach} & \textbf{EM} & \textbf{CM} & \textbf{ER} & \textbf{Mock Exec.} \\
\midrule
\multirow{4}{*}{API-only}
& Baseline & 0.00 & 20.31 & 0.00 & 28.45 \\
& RAG$_{Standard}$ & \textbf{4.21} & 45.23 & 44.41 & 71.58 \\
& RAG$_{RelevanceFiltering}$ & 4.15 & \textbf{61.26} & 44.06 & 68.51 \\
& RAG$_{IterativeRetrieval}$ & 3.51 & 61.14 & \textbf{54.55} & \textbf{79.30} \\
\midrule
\multirow{4}{*}{Mixed-doc.}
& Baseline & 0.00 & 20.31 & 0.00 & 28.45 \\
& RAG$_{Standard}$ & 3.86 & 57.09 & 42.66 & 75.79 \\
& RAG$_{RelevanceFiltering}$ & 3.48 & 44.74 & 39.16 & 68.64 \\
& RAG$_{IterativeRetrieval}$ & 3.51 & 36.45 & 29.82 & 50.53 \\
\bottomrule
\end{tabular}
\label{tab:api}
\end{table}

Under API-only settings, all retrieval configurations substantially improve performance. \textit{Standard RAG} achieves the highest exact match accuracy (4.21\%), while \textit{Relevance Filtering RAG} obtains the strongest component match accuracy (61.26\%). \textit{Iterative Retrieval RAG} achieves the highest endpoint retrieval accuracy (54.55\%) and mock execution success (79.30\%) in this setting.

Performance changes noticeably under mixed-documentation settings, where both database schemas and API documentation are available within the same retrieval environment. \textit{Standard RAG} maintains relatively stable mock execution success (75.79\%), whereas \textit{Relevance Filtering RAG} and \textit{Iterative Retrieval RAG} exhibit larger performance decreases. Component match accuracy for \textit{Relevance Filtering RAG} decreases from 61.26\% to 44.74\%, while mock execution success for \textit{Iterative Retrieval RAG} decreases from 79.30\% to 50.53\%. These results suggest that retrieval-control mechanisms may become sensitive to retrieval noise when semantically related but task-irrelevant documentation is retrieved together.

Paired two-tailed $t$-tests on exact match accuracy show no statistically significant differences between retrieval configurations in either API-only ($p \geq 0.18$) or mixed-documentation settings ($p \geq 0.70$). The corresponding 95\% confidence intervals for paired mean differences include zero, indicating that the observed differences are not statistically reliable. Overall, REST API call generation remains challenging across all evaluated configurations.

\subsection{SQL Query Generation}

Table \ref{tab:sql} presents the results for SQL query generation. Without retrieval augmentation, the baseline achieves 0\% exact match accuracy and 0\% mock execution success despite reaching 58.51\% component match accuracy. This indicates that the LLM possesses general SQL syntax knowledge but cannot reliably generate executable queries without access to the underlying database schema.

\begin{table}[h!]
\centering
\footnotesize
\setlength{\tabcolsep}{2.5pt}
\caption{SQL query generation results (\%).}
\begin{tabular}{llrrr}
\toprule
\textbf{Setting} & \textbf{Approach} & \textbf{EM} & \textbf{CM} & \textbf{Mock Exec.} \\
\midrule
\multirow{4}{*}{DB-only}
& Baseline & 0.00 & 58.51 & 0.00 \\
& RAG$_{Standard}$ & 14.45 & 78.34 & 71.97 \\
& RAG$_{RelevanceFiltering}$ & \textbf{15.61} & 78.24 & 72.83 \\
& RAG$_{IterativeRetrieval}$ & \textbf{15.61} & 77.85 & \textbf{73.12} \\
\midrule
\multirow{4}{*}{Mixed-doc.}
& Baseline & 0.00 & 58.51 & 0.00 \\
& RAG$_{Standard}$ & 11.56 & 76.59 & 69.08 \\
& RAG$_{RelevanceFiltering}$ & 10.98 & \textbf{78.86} & 67.63 \\
& RAG$_{IterativeRetrieval}$ & 15.32 & 78.22 & 70.81 \\
\bottomrule
\end{tabular}
\label{tab:sql}
\end{table}

Under database-only settings, all retrieval configurations achieve similar performance levels. Exact match accuracy ranges from 14.45\% for \textit{Standard RAG} to 15.61\% for both \textit{Relevance Filtering RAG} and \textit{Iterative Retrieval RAG}, while mock execution success remains near 72\%--73\%. These results suggest that retrieval strategy has limited influence when retrieval is restricted to homogeneous schema documentation.
The largest differences emerge under mixed-documentation settings. \textit{Standard RAG} decreases from 14.45\% to 11.56\% exact match accuracy, while \textit{Relevance Filtering RAG} decreases from 15.61\% to 10.98\%. In contrast, \textit{Iterative Retrieval RAG} maintains 15.32\% exact match accuracy, close to its 15.61\% under database-only retrieval.

These differences are statistically significant. Compared with \textit{Relevance Filtering RAG}, \textit{Iterative Retrieval RAG} increases exact match accuracy from 10.98\% to 15.32\%, corresponding to a 39.5\% relative improvement ($p = 0.0026 < 0.05$, mean difference = -0.04, 95\% CI [-0.07, -0.02]). Compared with \textit{Standard RAG}, exact match accuracy increases from 11.56\% to 15.32\%, corresponding to a 32.5\% relative improvement ($p = 0.0091 < 0.05$, mean difference = -0.04, 95\% CI [-0.07, -0.01]). Mean differences are reported as the comparator minus \textit{Iterative Retrieval RAG}, explaining the negative signs. No statistically significant differences are observed under database-only settings ($p \geq 0.15$).

The gap between exact match accuracy and mock execution success reflects the strictness of single-reference evaluation. Many generated SQL queries execute successfully despite failing exact match. However, successful execution alone does not establish semantic equivalence with the reference query. This observation is consistent with prior text-to-SQL evaluation studies~\cite{pourreza2023evaluating}.

\subsection{Combined Task}

Table \ref{tab:combined} presents results for the combined task, where the system must first route a request to SQL query generation or REST API call generation before producing the final output. Classification accuracy remains high across all evaluated configurations (95.89\%--97.79\%). However, this result reflects a deliberately constrained routing setting, as SQL requests are limited to \texttt{SELECT} operations while API requests exclude \texttt{GET} operations.

\begin{table}[h!]
\centering
\footnotesize
\setlength{\tabcolsep}{2.5pt}
\caption{Combined task results (\%) under mixed-documentation settings.}
\begin{tabular}{lrrrr}
\toprule
\textbf{Approach} & \textbf{EM} & \textbf{CM} & \textbf{Mock Exec.} & \textbf{Class} \\
\midrule
Baseline & 0.00 & 42.06 & 5.30 & \textbf{97.79} \\
RAG$_{Standard}$ & 7.45 & 67.68 & 67.10 & 97.00 \\
RAG$_{RelevanceFiltering}$ & 8.54 & 67.99 & 64.20 & 97.63 \\
RAG$_{IterativeRetrieval}$ & \textbf{10.29} & \textbf{70.77} & \textbf{68.93} & 95.89 \\
\bottomrule
\end{tabular}
\label{tab:combined}
\end{table}

Among the evaluated retrieval configurations, \textit{Iterative Retrieval RAG} achieves the highest exact match accuracy (10.29\%), component match accuracy (70.77\%), and mock execution success (68.93\%). Compared with \textit{Standard RAG}, exact match accuracy increases from 7.45\% to 10.29\%, corresponding to a 38.1\% relative improvement ($p = 0.0006 < 0.05$, mean difference = -0.03, 95\% CI [-0.05, -0.01]). Compared with \textit{Relevance Filtering RAG}, exact match accuracy increases from 8.54\% to 10.29\%, corresponding to a 20.5\% relative improvement ($p = 0.0283 < 0.05$, mean difference = -0.02, 95\% CI [-0.04, -0.00]). No statistically significant difference is observed between \textit{Standard RAG} and \textit{Relevance Filtering RAG} ($p = 0.1337$).

A task-level breakdown shows that the higher combined exact match accuracy of \textit{Iterative Retrieval RAG} is primarily associated with SQL generation under mixed documentation. For SQL-related requests, it achieves 15.90\% exact match accuracy versus 11.56\% for \textit{Standard RAG}, while the corresponding REST API differences are smaller.

Taken together, these results indicate task- and metric-dependent effects of iterative retrieval: \textit{Iterative Retrieval RAG} achieves higher exact match accuracy for SQL generation under mixed documentation, while REST API generation shows substantially lower mock execution success than \textit{Standard RAG} in the same documentation setting.

\section{Discussion}
\label{sec:discussion}

The results show that retrieval-control strategy affects structured generation performance when SQL query generation and REST API call generation are evaluated within shared retrieval environments. In particular, the observed effects vary across task types, evaluation metrics, and homogeneous versus mixed-documentation settings.

\subsection{Role of Retrieval Augmentation in Enterprise Structured Generation}

The most consistent observation across the experiments is the importance of retrieval augmentation in the evaluated setting. Without retrieval augmentation, the baseline achieves 0\% exact match accuracy across SQL query generation, REST API call generation, and the combined task. This indicates that GPT-5 alone was insufficient to match the task-specific reference outputs when generation depended on enterprise-specific schemas, endpoint definitions, and domain terminology. In contrast to general programming tasks that primarily rely on algorithmic reasoning, enterprise-oriented structured generation can require grounding in system-specific knowledge and documentation \cite{idrisov2024program}.

Introducing retrieval augmentation also substantially improves mock execution success across the evaluated tasks. Under single-documentation settings, mock execution success exceeds 70\% for both SQL query generation and REST API call generation. These results are consistent with prior work showing that documentation retrieval improves code generation performance for unfamiliar APIs and libraries \cite{zhou2023docprompting,chen2025llms,kivroglou2025investigating}. Our results show similar benefits in the evaluated enterprise-oriented setting involving database schemas and REST API specifications.

\subsection{Retrieval Strategy Sensitivity Under Mixed-Documentation Settings}

The experiments further show that retrieval behavior changes when database schemas and API documentation are available within the same retrieval environment. Under database-only settings, all retrieval configurations achieve comparable SQL query generation performance. Under mixed-documentation settings, however, performance differences emerge across configurations and metrics.

\textit{Iterative Retrieval RAG} maintains nearly identical SQL exact match accuracy under mixed-documentation settings (15.32\%) compared with database-only settings (15.61\%), while \textit{Standard RAG} and \textit{Relevance Filtering RAG} both exhibit decreases. This pattern suggests that iterative retrieval through query decomposition may be less sensitive to mixed-documentation retrieval for SQL exact match accuracy in the evaluated setting.

\textit{Relevance Filtering RAG} also shows performance decreases under mixed-documentation retrieval, particularly in REST API component match accuracy and SQL exact match accuracy. One possible explanation is that relevance filtering may discard useful information when semantically related documentation from different sources is retrieved simultaneously. \textit{Standard RAG} shows a different pattern, with comparatively stable REST API mock execution success but lower SQL exact match accuracy under mixed documentation.

These findings suggest that retrieval-control effectiveness depends on the task and metric, consistent with broader evidence that RAG performance varies across reasoning domains \cite{krazius2026crossdomain}.

\subsection{Differences Between SQL Query Generation and REST API Call Generation}

The experiments also reveal notable differences between SQL query generation and REST API call generation. Under mixed-documentation settings, \textit{Iterative Retrieval RAG} achieves higher exact match accuracy for SQL query generation, whereas REST API call generation remains comparatively difficult across all retrieval configurations.

One possible explanation is that SQL generation often depends on combining multiple schema elements distributed across documentation chunks, including table relationships, column definitions, and filtering constraints. Iterative retrieval may therefore help gather complementary schema information across multiple retrieval steps. However, the present evaluation does not directly measure retrieval quality or context composition, so this interpretation should be considered a possible explanation rather than a demonstrated mechanism.

REST API call generation, in contrast, depends on selecting the correct endpoint and constructing valid parameter structures. In the evaluated setting, additional retrieval iterations do not consistently improve performance across metrics. In particular, \textit{Iterative Retrieval RAG} achieves lower mock execution success than \textit{Standard RAG} under mixed documentation, despite competitive results under API-only retrieval.

Mock execution success for REST API call generation remains substantially higher than exact match accuracy. This shows that outputs can execute successfully in the mock environment despite differing from the reference output, but execution alone does not establish semantic correctness. A similar distinction applies to SQL query generation, where alternative formulations may differ structurally from the reference query. This limitation has also been discussed in prior text-to-SQL evaluation work \cite{pourreza2023evaluating,yu2019spider}.

\subsection{Implications and Deployment Recommendations}

The findings have several implications for enterprise-oriented LLM systems. First, retrieval plays an important role when structured generation depends on system-specific schemas and APIs. In our evaluation, exact match accuracy remained at 0\% across all tasks without retrieval.

Second, retrieval strategy selection can become more important when heterogeneous documentation sources are combined. Under mixed-documentation settings, \textit{Iterative Retrieval RAG} maintains higher SQL exact match accuracy than \textit{Standard RAG} and \textit{Relevance Filtering RAG}, while REST API results show a different pattern.

Third, the combined-task results show that routing between SQL query generation and REST API call generation can be integrated with structured generation. Classification accuracy is high in the evaluated setting, although the routing task is constrained by the separation between \texttt{SELECT}-based SQL requests and non-\texttt{GET} API operations.

Finally, mock execution checks can complement structural evaluation by showing whether generated outputs execute in the controlled environment. However, execution success does not establish semantic correctness and should therefore be combined with structural or semantic validation.

\subsection{Limitations}

Several limitations affect the generalizability of this study: First, all experiments use GPT-5 as a fixed backbone, and GPT-5 is also used to generate and humanize candidate test cases. Although all candidates subsequently undergo mock execution screening and independent expert review, future work should assess whether the observed retrieval patterns generalize across other LLM architectures and independently constructed benchmarks. Second, the evaluation is restricted to a single enterprise domain and a dataset of 631 cases. Third, all configurations are evaluated in a single run using the default API decoding configuration; repeated runs would allow assessment of stochastic variation. Fourth, the combined routing task is constrained because SQL requests are limited to \texttt{SELECT}, whereas API requests exclude \texttt{GET}, and therefore does not represent more ambiguous routing scenarios.

In addition, evaluation relies on mock execution environments and single-reference outputs. Mock execution success captures whether an output executes in the controlled environment but does not establish semantic correctness or equivalence with the reference. Production environments may also introduce constraints such as authentication, transactional consistency, or network failures not represented here. Finally, the retrieval configurations do not use matched computational budgets: \textit{Iterative Retrieval RAG} may perform multiple retrieval and LLM steps, whereas \textit{Standard RAG} uses a single retrieval and generation step. The study therefore compares configurations as implemented rather than under equal latency, token, or computational budgets.

Despite these limitations, the controlled evaluation setup enables systematic comparison of the implemented retrieval configurations under different documentation conditions and provides evidence of task- and metric-dependent behavior in shared enterprise retrieval environments.

\section{Conclusion and Future Work}
\label{sec:conclusion}

\subsection{Conclusion}

This paper presented an evaluation of retrieval-augmented generation configurations for SQL query generation and REST API call generation within a shared enterprise retrieval environment. Using SAP Transactional Banking documentation, we compared \textit{Standard RAG}, \textit{Relevance Filtering RAG}, and \textit{Iterative Retrieval RAG} across database-only, API-only, and mixed-documentation settings.

The results show that retrieval augmentation substantially improves structured generation performance in the evaluated setting. Without retrieval augmentation, exact match accuracy remained at 0\% across all tasks, whereas retrieval achieved mock execution success of up to 79.30\% and component match accuracy of up to 78.86\%.

The experiments show task- and metric-dependent behavior under mixed-documentation settings. \textit{Iterative Retrieval RAG} achieves higher SQL exact match accuracy than the other configurations, but substantially lower REST API mock execution success than \textit{Standard RAG}. Thus, no retrieval configuration is uniformly superior across tasks and metrics.

At the same time, REST API call generation remains challenging across all evaluated configurations, particularly with respect to exact match accuracy. The differences between exact match accuracy and mock execution success further show that successful execution in the controlled mock environment captures a different property than structural agreement and does not by itself establish semantic correctness.

Overall, the study shows that evaluating retrieval strategies in shared enterprise environments can reveal differences not visible in isolated SQL-only or API-only benchmarks \cite{yu2019spider,kim2024seal}. The findings further indicate that retrieval-control configurations should be evaluated for the specific task, metric, and documentation setting rather than assuming a general advantage of one strategy.

\subsection{Future Work}

Several directions may extend the present work:

First, future studies should evaluate retrieval behavior across additional LLM architectures, including specialized coding-oriented models, open-source alternatives, and collaborative multi-agent retrieval architectures for enterprise structured generation \cite{idrisov2025multiagent}. Repeated runs should additionally be used to quantify stochastic variation across configurations.

Second, the evaluation should extend to additional enterprise domains with different documentation structures, schema complexities, and API conventions to improve generalizability.

Third, future work should investigate multi-step workflows involving sequential combinations of SQL queries and REST API calls, as well as more ambiguous routing settings in which the target interface cannot be inferred primarily from the distinction between read and transactional operations. Such settings would more closely reflect realistic enterprise assistants in which information retrieval and transactional operations interact dynamically.

Fourth, retrieval quality may benefit from alternative retrieval and chunking strategies, including hybrid dense-sparse retrieval, hierarchical retrieval, or query-adaptive chunking approaches specifically designed for structured enterprise documentation. Future comparisons should also consider matched retrieval and computational budgets and report latency and cost alongside generation quality.

Finally, future evaluation should further investigate semantic equivalence and multi-reference evaluation. Both SQL queries and REST API calls may admit multiple correct realizations, suggesting that mock execution should complement, rather than replace, structural and semantic evaluation \cite{yu2019spider,pourreza2023evaluating}.

\bibliographystyle{IEEEtran}
\bibliography{references.bib}

@article{koutrika2024natural,
  title={{Natural Language Data Interfaces: A Data Access Odyssey}},
  author={Koutrika, Georgia},
  journal={LIPIcs},
  volume={ICDT 2024},
  year={2024},
  doi={10.4230/LIPIcs.ICDT.2024.1}
}

@inproceedings{wang2024fantastic,
  title={{FANTAstic SEquences and Where to Find Them: Faithful and Efficient API Call Generation Through State-Tracked Constrained Decoding and Reranking}},
  author={Wang, Zhuoer and Ribeiro, Leonardo F. R. and Papangelis, Alexandros and Mukherjee, Rohan and Wang, Tzu-Yen and Zhao, Xinyan and Biswas, Arijit and Caverlee, James and Metallinou, Angeliki},
  booktitle={EMNLP 2024},
  year={2024},
  doi={10.18653/v1/2024.findings-emnlp.359}
}

@article{usta2021dbtagger,
  title={{DBTagger: Multi-Task Learning for Keyword Mapping in NLIDBs Using Bi-Directional Recurrent Neural Networks}},
  author={Usta, Arif and Karakayali, Akifhan and Ulusoy, {\"O}zg{\"u}r},
  journal={VLDB Endowment},
  volume={14},
  number={5},
  pages={813--821},
  year={2021},
  doi={10.14778/3446095.3446103}
}

@article{lennerholt2020user,
  title={{User-Related Challenges of Self-Service Business Intelligence}},
  author={Lennerholt, Carl and Van Laere, Jorgen and S{\"o}derstr{\"o}m, Eva},
  journal={Information Systems Management},
  volume={38},
  number={4},
  pages={309--323},
  year={2020},
  doi={10.1080/10580530.2020.1814458}
}

@article{chen2021evaluating,
  title={{Evaluating Large Language Models Trained on Code}},
  author={Chen, Mark and Tworek, Jerry and Jun, Heewoo and Yuan, Qiming and de Oliveira Pinto, Henrique Ponde and Kaplan, Jared and Edwards, Harri and Burda, Yuri and Joseph, Nicholas and Brockman, Greg and others},
  journal={arXiv:2107.03374},
  year={2021}
}

@article{huynh2025survey,
  title={{A Survey on Large Language Models for Code Generation}},
  author={Huynh, Nhi and Lin, Bill},
  journal={arXiv:2503.01245},
  year={2025}
}

@article{chen2025llms,
  title={{When LLMs Meet API Documentation: Can Retrieval Augmentation Aid Code Generation Just as It Helps Developers?}},
  author={Chen, Junda and Chen, Shuzheng and Cao, Jie and Shen, Jing and Cheung, Shing-Chi},
  journal={arXiv:2503.15231},
  year={2025}
}

@inproceedings{zhou2023docprompting,
  title={{DocPrompting: Generating Code by Retrieving the Docs}},
  author={Zhou, Shuyan and Alon, Uri and Xu, Frank F. and Wang, Zhiruo and Jiang, Zhengbao and Neubig, Graham},
  booktitle={ICLR 2023},
  year={2023}
}

@article{lewis2020retrieval,
  title={{Retrieval-Augmented Generation for Knowledge-Intensive NLP Tasks}},
  author={Lewis, Patrick and Perez, Ethan and Piktus, Aleksandra and Petroni, Fabio and Karpukhin, Vladimir and Goyal, Naman and K{\"u}ttler, Heinrich and Lewis, Mike and Yih, Wen-tau and Rockt{\"a}schel, Tim and Riedel, Sebastian and Kiela, Douwe},
  journal={NeurIPS 2020},
  year={2020}
}

@inproceedings{yu2019spider,
  title={{Spider: A Large-Scale Human-Labeled Dataset for Complex and Cross-Domain Semantic Parsing and Text-to-SQL Task}},
  author={Yu, Tao and Zhang, Rui and Yang, Kai and Yasunaga, Michihiro and Wang, Dongxu and Li, Zifan and Ma, James and Li, Irene and Yao, Qingning and Roman, Shanelle and others},
  booktitle={EMNLP 2018},
  year={2018},
  address={Brussels, Belgium},
  doi={10.18653/v1/D18-1425}
}

@article{kim2024seal,
  title={{SEAL: Suite for Evaluating API-Use of LLMs}},
  author={Kim, Wonyoung and Jagmohan, Ashish and Vempaty, Aditya},
  journal={arXiv:2409.15523},
  year={2024}
}

@inproceedings{asai2024selfrag,
  title={{Self-RAG: Learning to Retrieve, Generate, and Critique Through Self-Reflection}},
  author={Asai, Akari and Wu, Zeqiu and Wang, Yizhong and Sil, Avirup and Hajishirzi, Hannaneh},
  booktitle={International Conference on Learning Representations (ICLR)},
  year={2024}
}

@article{wang2025chain,
  title={{Chain-of-Retrieval Augmented Generation}},
  author={Wang, Liang and Chen, Haoyang and Yang, Nan and Huang, Xueguang and Dou, Zhicheng and Wei, Furu},
  journal={arXiv:2501.14342},
  year={2025}
}

@inproceedings{liu2024apigen,
  title={{APIGen: Automated Pipeline for Generating Verifiable and Diverse Function-Calling Datasets}},
  author={Liu, Zuxin and Hoang, Thai and Zhang, Jianguo and Zhu, Ming and Lan, Tian and Kokane, Shirley and Tan, Juntao and Yao, Weiran and Liu, Zhiwei and Feng, Yihao and Murthy, Rithesh and Yang, Liangwei and Savarese, Silvio and Niebles, Juan Carlos and Wang, Huan and Heinecke, Shelby and Xiong, Caiming},
  booktitle={NeurIPS 2024},
  year={2024},
  doi={10.52202/079017-1725}
}

@inproceedings{yu2019cosql,
  title={{CoSQL: A Conversational Text-to-SQL Challenge Towards Cross-Domain Natural Language Interfaces to Databases}},
  author={Yu, Tao and Zhang, Rui and Er, Heyang Yasunaga and Li, Suyi and Xue, Eric and Pang, Bo and Lin, Xi Victoria and Tan, Yi Chern and Shi, Tianze and Li, Zihan and others},
  booktitle={EMNLP-IJCNLP 2019},
  year={2019},
  address={Hong Kong, China},
  doi={10.18653/v1/D19-1204}
}

@inproceedings{yu2019sparc,
  title={{SParC: Cross-Domain Semantic Parsing in Context}},
  author={Yu, Tao and Zhang, Rui and Yasunaga, Michihiro and Tan, Yi Chern and Lin, Xi Victoria and Li, Suyi and Er, Heyang and Li, Irene and Pang, Bo and Chen, Tao and others},
  booktitle={ACL 2019},
  pages={4511--4523},
  year={2019},
  address={Florence, Italy},
  publisher={Association for Computational Linguistics},
  doi={10.18653/v1/P19-1443}
}

@inproceedings{lei2024evaluating,
  title={{Spider 2.0: Evaluating Language Models on Real-World Enterprise Text-to-SQL Workflows}},
  author={Lei, Fangyu and Chen, Jixuan and Ye, Yuxiao and Cao, Ruisheng and Shin, Dongchan and Su, Hongjin and Suo, Zhaoqing and Gao, Hongcheng and Hu, Wenjing and Yin, Pengcheng and Zhong, Victor and Xiong, Caiming and Sun, Ruoxi and Liu, Qian and Wang, Sida and Yu, Tao},
  booktitle={ICLR 2025},
  year={2025}
}

@inproceedings{pourreza2023evaluating,
  title={{Evaluating Cross-Domain Text-to-SQL Models and Benchmarks}},
  author={Pourreza, Mohammadreza and Rafiei, Davood},
  booktitle={EMNLP 2023},
  year={2023},
  doi={10.18653/v1/2023.emnlp-main.99}
}

@inproceedings{schick2023toolformer,
  title={{Toolformer: Language Models Can Teach Themselves to Use Tools}},
  author={Schick, Timo and Dwivedi-Yu, Jane and Dess{\`\i}, Roberto and Raileanu, Roberta and Lomeli, Maria and Zettlemoyer, Luke and Cancedda, Nicola and Scialom, Thomas},
  booktitle={NeurIPS 2023},
  year={2023}
}

@inproceedings{li2023apibank,
  title={{API-Bank: A Comprehensive Benchmark for Tool-Augmented LLMs}},
  author={Li, Minghao and Zhao, Yingxiu and Yu, Bowen and Song, Feifan and Li, Hangyu and Yu, Haiyang and Li, Zhoujun and Huang, Fei and Li, Yongbin},
  booktitle={EMNLP 2023},
  year={2023},
  doi={10.18653/v1/2023.emnlp-main.187}
}

@inproceedings{qin2023tool,
  title={{ToolLLM: Facilitating Large Language Models to Master 16000+ Real-World APIs}},
  author={Qin, Yujia and Liang, Shihao and Ye, Yining and Zhu, Kunlun and Yan, Lan and Lu, Yaxi and Lin, Yankai and Cong, Xin and Tang, Xiangru and Qian, Bill and Zhao, Sihan and Hong, Lauren and Tian, Runchu and Xie, Ruobing and Zhou, Jie and Gerstein, Mark and Li, Dahai and Liu, Zhiyuan and Sun, Maosong},
  booktitle={ICLR 2024},
  year={2024}
}

@inproceedings{patil2023gorilla,
  title={{Gorilla: Large Language Model Connected with Massive APIs}},
  author={Patil, Shishir G. and Zhang, Tianjun and Wang, Xin and Gonzalez, Joseph E.},
  booktitle={NeurIPS 2024},
  year={2024},
  doi={10.52202/079017-4020}
}

@inproceedings{basu2025nestful,
  title={{NESTFUL: A Benchmark for Evaluating LLMs on Nested Sequences of API Calls}},
  author={Basu, Kinjal and Abdelaziz, Ibrahim and Kate, Kiran and Agarwal, Mayank and Crouse, Maxwell and Rizk, Yara and Bradford, Kelsey and Munawar, Asim and Kumaravel, Sadhana and Goyal, Saurabh and Wang, Xin and Lastras, Luis A. and Kapanipathi, Pavan},
  booktitle={EMNLP 2025},
  year={2025},
  doi={10.18653/v1/2025.emnlp-main.1702}
}

@inproceedings{jain2024livecode,
  title={{LiveCodeBench: Holistic and Contamination Free Evaluation of Large Language Models for Code}},
  author={Jain, Naman and Han, King and Gu, Alex and Li, Wen-Ding and Yan, Fanjia and Zhang, Tianjun and Wang, Sida and Solar-Lezama, Armando and Sen, Koushik and Stoica, Ion},
  booktitle={ICLR 2025},
  year={2025}
}

@misc{sap2025trbk,
  title={{SAP Transactional Banking API Documentation}},
  author={{SAP SE}},
  year={2025},
  howpublished={\url{https://api.sap.com/package/TRBK/odatav4}},
  note={Accessed: 2025-05-21}
}

@misc{postman2025,
  title={{Postman Mock Servers Documentation}},
  author={{Postman}},
  year={2025},
  howpublished={\url{https://learning.postman.com/docs/design-apis/mock-apis/overview/}},
  note={Accessed: 2025-05-21}
}

@article{amiri2024chunk,
  title={{Chunk Twice, Embed Once: A Systematic Study of Segmentation and Representation Trade-offs in Chemistry-Aware Retrieval-Augmented Generation}},
  author={Amiri, Mahsa and Bocklitz, Thomas},
  journal={arXiv:2506.17277},
  year={2025}
}

@article{bhat2025rethinking,
  title={{Rethinking Chunk Size for Long-Document Retrieval: A Multi-Dataset Analysis}},
  author={Bhat, Sai Rohith and Rudat, Moritz and Spiekermann, Julius and Flores-Herr, Nick},
  journal={arXiv:2505.21700},
  year={2025}
}

@inproceedings{kivroglou2025investigating,
  title={{Investigating Retrieval Augmented Generation for LLM-Based Code Generation}},
  author={Kivroglou, Paraskevi and Schlippe, Tim and Martin, Simon},
  booktitle={FLLM 2025},
  address={Vienna, Austria},
  year={2025},
  month={November}
}

@article{idrisov2024program,
  title={{Program Code Generation with Generative AIs}},
  author={Idrisov, Baskhad and Schlippe, Tim},
  journal={Algorithms},
  volume={17},
  number={2},
  pages={62},
  year={2024},
  month={January},
  publisher={MDPI},
  note={Special Issue: Feature Papers on Artificial Intelligence Algorithms and Their Applications},
  doi={10.3390/a17020062}
}

@inproceedings{idrisov2025multiagent,
  title={{Program Code Generation: Single LLMs vs. Multi-Agent Systems}},
  author={Idrisov, Baskhad and Eisenacher, Esther and Schlippe, Tim},
  booktitle={ICNLP 2025},
  address={Guangzhou, China},
  year={2025},
  month={March}
}

@inproceedings{ryser2025calibrated,
  title={{Calibrated Trust in Dealing with LLM Hallucinations: A Qualitative Study}},
  author={Ryser, Adrian and Allwein, Florian and Schlippe, Tim},
  booktitle={FLLM 2025},
  address={Vienna, Austria},
  year={2025},
  month={November}
}

@inproceedings{krazius2026crossdomain,
  title={{A Cross-Domain Evaluation of Prompting Techniques for Large Language Models}},
  author={Krazius, Kevin and Schlippe, Tim and Libbrecht, Paul},
  booktitle={FLLM 2026},
  year={2026}
}

\end{document}